# Brain experiments imply adaptation mechanisms which outperform common AI learning algorithms

Shira Sardi[1], Roni Vardi[2], Yuval Meir[1], Yael Tugendhaft[1], Shiri Hodassman[1], Amir Goldental[1] & Ido Kanter[1,2,*]

[1]Department of Physics, Bar-Ilan University, Ramat-Gan, 52900, Israel.

[2]Gonda Interdisciplinary Brain Research Center and the Goodman Faculty of Life Sciences, Bar-Ilan University, Ramat-Gan, 52900, Israel.

[*]e-mail: ido.kanter@biu.ac.il

**Attempting to imitate the brain's functionalities, researchers have bridged between neuroscience and artificial intelligence for decades; however, experimental neuroscience has not directly advanced the field of machine learning (ML). Here, using neuronal cultures, we demonstrate that increased training frequency accelerates the neuronal adaptation processes. This mechanism was implemented on artificial neural networks, where a local learning step-size increases for coherent consecutive learning steps, and tested on a simple dataset of handwritten digits, MNIST. Based on our on-line learning results with a few handwriting examples, success rates for brain-inspired algorithms substantially outperform the commonly used ML algorithms. We speculate this emerging bridge from slow brain function to ML will promote ultrafast decision making under limited examples, which is the reality in many aspects of human activity, robotic control, and network optimization.**

## Introduction

Machine learning is based on Donald Hebb's pioneering work; seventy years ago, he suggested that learning occurs in the brain through synaptic (link) strength modifications (1). A synaptic strength modification typically lasts tens of minutes (2) while the clock speed of a neuron (node) ranges around one second (3). Although the brain is comparatively slow, its computational capabilities outperform typical state-of-the-art artificial intelligence algorithms. Following this speed/capability paradox, we experimentally derive accelerated learning mechanisms based on small datasets, where their utilization on gigahertz processors is expected to lead to ultrafast decision making.

Unlike modern computers, a well-defined global clock does not govern brain dynamics; instead, they are a function of relative event timing (e.g., stimulations and evoked spikes) (4). According to neuronal computational, using decaying input summation via its ramified dendritic trees, each neuron sums the asynchronous incoming electrical signals and generates a short electrical pulse (spike) when its threshold is reached. For each neuron, synaptic strength is slowly modified based on the relative timing of inputs from other synapses; if a signal is induced from a synapse without generating a spike, its associated strength is modified based on the relative timing to adjacent spikes from other synapses on the same neuron (5).

Recently it was experimentally demonstrated that each neuron functions as a collection of independent threshold units (6). After signals arrive via one of the dendritic trees, each threshold unit is activated. Additionally, a new type of adaptive rule was experimentally observed based on dendritic signal arrival timing (7), which is similar to the slow adaptation mechanism currently attributed to synapses (links). This dendritic adaptation occurs on a faster timescale: it requires approximately five minutes, while synaptic modification requires tens of minutes or more.

## Results

In this study, dendritic adaptation was experimentally examined at a higher stimulation frequency, 5 Hz, using the training pattern of previous experiments run at 0.5 to 1 Hz. We planted neuronal cultures on a multi-electrode-array with added synaptic blockers, which extracellularly stimulated a patched neuron via its dendrites (Fig. 1a and Materials and Methods). The adaptation process consisted of a training set: 50 pairs of stimulations. After an above-threshold intracellular stimulation, an extracellular stimulation that did not evoke a spike arrived with a predefined delay, typically 1 to 4 ms (Fig. 1b). We primarily take into account differing extra- and intra-spike waveforms (Fig. 1a, right), which presumably activated the neuron from two independent dendritic trees (7).

By comparing the amplitudes of intracellular responses and extracellular stimulation before and after the training procedure, we quantified the effect of the neuronal adaptation. To quantify the initial response, the extracellular stimulation amplitude was decreased until no reliable evoked spikes were observed (Fig. 1, c and d, left).

In the first type of experiment, one minute after the training terminated, we measured the enhanced responses and witnessed dendritic adaptation (Fig. 1c, right). When compared with the visible adaptation time for the 1 Hz training, that of the 5 Hz training was substantially faster. Occasionally, the visible effect of adaptation was further enhanced after more time passed (Supplementary Fig. 1), suggesting by extrapolation that adaptation might occur in much less than one minute. Because after training, the initialization and compilation of subsequent experiments required a minimal time lag (one minute), the feasibility of such ultrafast adaptation was impossible to examine.

To overcome this limitation, we introduced a second type of experiment (Fig. 1d): shortly after the end of the training procedure, the neuron was extracellularly stimulated using two predefined amplitudes with unreliable responses. Without requiring any new experimental methods or compilations (Materials and Methods), this procedure pinpointed dendritic adaptation within only 10 seconds of the training termination (Fig. 1d, right).

With the increased training frequency, the adaptation process substantially accelerated (Fig. 2a), potentially implying a time-dependent decaying adaptation step-size:

$$\eta_{adap}^{t+1} = \eta_{adap}^{t} \cdot e^{-\frac{\tau}{\tau_0}} + \Delta \qquad (1)$$

where the current adaptation step, $\eta_{adap}^{t+1}$, is equal to the previous one with a decaying weight, t stands for a discrete time step, $\tau_0$ is a constant, $1/\tau$ stands for the training frequency, and $\Delta$ is a constant representing the incremental effect of the current training step. This type of decay process occurs in many biological scenarios and represents, for instance, the decaying concentration of active material due to diffusion. As a generalization of Eq. (1), incoherent consecutive training steps are also allowed, decreasing the dendritic strength and resulting in a -$\Delta$ term in Eq. (1).

Using supervised on-line learning of realizable rules and binary classification (Fig. 2, b and c), we first examined the impact of the time-dependent adaptation steps (Eq. 1) on accelerating biological learning processes. The teacher provided the student asynchronous-input and binary-output relations (8), where both had the same architecture of the simplest classifier, the perceptron (9), and the output nodes were represented by a leaky integrate-and-fire neuron (10). Two scenarios were examined: synaptic adaptation and dendritic adaptation (Fig. 2, b and c and Materials and Methods). Results clearly indicate that the generalization error, $\varepsilon_g$, of the experimentally-inspired time-dependent η (Eq. 1) substantially outperformed the fixed η scenario (Fig. 2, b and c). This accelerated learning stems from the fact that weights in synaptic learning convergence to the extreme limits, vanishing or above-threshold weights (7). Hence, the learning step-sizes in coherent dynamics increased toward these extremes, accelerating the learning scenario (Fig. 2b). Similarly, in the dendritic case (Fig. 2c), weights oscillated and synchronized via repeatedly hitting the boundary values (7). Hence, a faster decay of $\varepsilon_g$ also resulted from learning step acceleration (Fig. 2c).

Next, we examined the experimentally-inspired time-dependent learning step mechanism on the supervised learning of an unrealizable rule using the MNIST database (11) tested on a neural network. This database consists of a large number of examples of handwritten digits (Fig. 3a) and is commonly used as a prototypical problem for quantifying the generalization performance of machine learning algorithms for various image processing tasks. In this study we use a small subset of the MNIST database without any data extension methods (12-14). The commonly used trained networks consisted of 784 inputs representing the 28 x 28 pixels of a digit, one hidden layer (30 units in this study), and ten outputs representing the labels (Fig. 3a). The commonly used learning approach is the backpropagation strategy (15):

$$W^{t+1} = W^t - \eta \cdot \nabla_{W^t} C \qquad (2)$$

where weight at time-step $t$, $W^t$, is modified with a step-size η towards the minus sign of the gradient of the cost function, C. An improved approach is the momentum strategy (5, 16, 17) and regularization of the weights (18, 19):

$$W^{t+1} = (1 - \alpha) \cdot W^t + V^{t+1} \qquad (3)$$
$$V^{t+1} = \mu \cdot V^t - \eta_0 \cdot \nabla_{W^t} C$$

where the momentum, μ, and the regularization, α, are constants in the region [0,1] and $\eta_0$ is a constant. We optimized the performance of the momentum strategy (Eq. 3) over $(\mu, \alpha, \eta_0)$ for a limited training dataset using the cross-entropy cost function (Materials and Methods) and compared its performance with the following two experimentally-inspired learning mechanisms consisting of time-dependent η.

In the first approach, acceleration, the time-dependent η, and the update rules for weight are given by these equations:

$$W^{t+1} = (1 - \alpha) \cdot W^t - |\eta^{t+1}| \cdot \nabla_{W^t} C \qquad (4)$$
$$\eta^{t+1} = \eta^t \cdot e^{-\tau} + A_{1/2} \cdot tanh(\beta_{1/2} \cdot \nabla_{W^t} C)$$

where τ is the positive decaying factor, $A_1$ and $\beta_1$ are constants representing the amplitude and the gain between the input and the hidden layers, respectively, and $A_2$ and $\beta_2$ represent the same between the hidden and the output layers. It is evident that coherent consecutive gradients of weight, i.e., with the same sign, increased its conjugate η. Note, in the limit $\beta \to \infty$, the equation for η was simplified, $\eta^{t+1} = \eta^t \cdot \exp(-\tau) + A \cdot sign(\nabla_{W^t} C)$. The second approach, advanced acceleration, combines the two previous approaches (Eq. 3,4):

$$W^{t+1} = (1 - \alpha) \cdot W^t + V^{t+1} \qquad (5)$$
$$V^{t+1} = \mu \cdot V^t - |\eta^{t+1}| \cdot \nabla_{W^t} C$$
$$\eta^{t+1} = \eta^t \cdot e^{-\tau} + A_{1/2} \cdot tanh(\beta_{1/2} \cdot \nabla_{W^t} C)$$

For the two experimentally-inspired accelerated-approaches (Eq. 4, 5), the changes in weights and η depend on the higher moments of gradients, in contrast with the linear dependence of the momentum approach (Eq. 3). Given a limited subset of the dataset examples the performance of the accelerated approaches was maximized over six (Eq. 4) and seven (Eq. 5) parameters (Materials and Methods).

The on-line training set consisted of 300 randomly chosen examples: each label appeared 30 times within a random order. After 300 learning steps, the accelerated approaches outperformed the momentum method by more than 25 %, and the test accuracy increased from about 0.43 to 0.54, respectively (Fig. 3b). Note, the acceleration approach (Eq. 4) showed similar performance to the advanced acceleration approach (Eq. 5). These improved results were found to be robust also for on-line training based on 6000 examples (30 batches of size 200) (Fig. 3c) and 1200 examples (24 batches of size 50) (Fig. 3d). For both cases (Fig. 3, c and d), the advanced acceleration approach (Eq. 5) offered the best performance. We repeated the training using the same 60 examples 5 times (60 x 5 = 300 training in total); compared with using 300 examples for training once, the 60 x 5 approach yielded better performance: using the advanced acceleration approach test accuracy increased from 0.54 to 0.57 (Fig. 3e and Supplementary Fig. 2). For a given number of network updates, results demonstrate that smaller example sets yield more information. This result stems from the random training order of a randomly selected small dataset, e.g., 60 or 300 examples, consisting of a balanced appearance for each label. Around equalized trained label appearances, there are more temporal fluctuations for a dataset involving 300 examples with 30 appearances for each label than for one involving 5 sets of 60 examples with 6 appearances for each label. Indeed, for a training set of 300 distinct examples composed of 5 subsets of 60 balanced examples, a test accuracy of 0.57 was achieved (Supplementary Fig. 3), and for one with 30 subsets of 10 examples where each label appears once, the test accuracy increased further to 0.67 and to 0.7 for a fixed label order (Supplementary Fig. 4). Results indicate that in order to maximize the test accuracy for on-line scenarios and especially for small datasets, the balanced set of examples and their balanced temporal training order are important ingredients.

**Conclusions**

Based on increased $\eta$ with coherent consecutive gradients, the brain-inspired accelerated-learning mechanism outperforms existing common ML strategies for small sets of training examples (*20*). Consistent results occur across various cost functions, e.g., square cost-function, however, with a relatively diminished performance (Fig. 3f). Because the performance maximization for a given dataset depends on the selected acceleration approach (Fig. 3, b and c), adapting the learning approach during the training process may improve performance. Nevertheless, in addition to possible advanced nonlinear functions for updating $\eta$, given the number of network updates, the ultimate scheduling of acceleration approaches and the ordering of trained examples to maximize the performance deserves further research. The presented bridge from experimental neuroscience to ML is expected to further advance decision making using limited databases, which is the reality in many aspects of human activity (*21*), robotic control (*22, 23*), and network optimization (*24, 25*).

**Materials and Methods**

**In-Vitro Experiments:**

Animals:

All procedures were in accordance with the National Institutes of Health Guide for the Care and Use of Laboratory Animals and Bar-Ilan University Guidelines for the Use and Care of Laboratory Animals in Research and were approved and supervised by the Bar-Ilan University Animal Care and Use Committee.

Culture preparation:

Cortical neurons were obtained from newborn rats (Sprague-Dawley) within 48 hours after birth using mechanical and enzymatic procedures. The cortical tissue was digested enzymatically with 0.05% trypsin solution in phosphate-buffered saline (Dulbecco's PBS) free of calcium and magnesium, and supplemented with 20 mM glucose, at 37°C. Enzyme treatment was terminated using heat-inactivated horse serum, and cells were then mechanically dissociated mostly by trituration. The neurons were plated directly onto substrate-integrated multi-electrode arrays (MEAs) and allowed to develop functionally and structurally mature networks over a time period of 2-4 weeks in-vitro, prior to the experiments. The number of plated neurons in a typical network was in the order of 1,300,000, covering an area of about ~5 cm$^2$. The preparations were bathed in minimal essential medium (MEM-Earle, Earle's Salt Base without L-Glutamine) supplemented with heat-inactivated horse serum (5%), B27 supplement (2%), glutamine (0.5 mM), glucose (20 mM), and gentamicin (10 g/ml), and maintained in an atmosphere of 37°C, 5% $CO_2$ and 95% air in an incubator.

Synaptic blockers:

Experiments were conducted on cultured cortical neurons that were functionally isolated from their network by a pharmacological block of glutamatergic and GABAergic synapses. For each culture at least 20 μl of a cocktail of synaptic blockers were used, consisting of 10 μM CNQX (6-cyano-7-nitroquinoxaline-2,3-dione), 80 μM APV (DL-2-amino-5-phosphonovaleric acid) and 5 μM Bicuculline methiodide. After this procedure no spontaneous activity was observed both in the MEA and the patch clamp recordings. In addition, repeated extracellular stimulations did not provoke the slightest cascades of neuronal responses (recorded extra- or intra- cellular).

Stimulation and recording – MEA:

An array of 60 Ti/Au/TiN extracellular electrodes, 30 μm in diameter, and typically spaced 200 μm from each other (Multi-Channel Systems, Reutlingen, Germany) was used. The insulation layer (silicon nitride) was pre-treated with polyethyleneimine (0.01% in 0.1 M Borate buffer solution). A commercial setup (MEA2100-60-headstage, MEA2100-interface board, MCS, Reutlingen, Germany) for recording and analyzing data from 60-electrode MEAs was used, with integrated data acquisition from 60 MEA electrodes and 4 additional analog channels, integrated filter amplifier and 3-channel current or voltage stimulus generator. Each channel was sampled at a frequency of 50k samples/s, thus the recorded action potentials and the changes in the neuronal response latency were measured at a resolution of 20 μs. Mono-phasic square voltage pulses were used, in the range of [−900, −100] mV and [100, 2000] μs.

Stimulation and recording – Patch Clamp:

The Electrophysiological recordings were performed in whole cell configuration utilizing a Multiclamp 700B patch clamp amplifier (Molecular Devices, Foster City, CA). The cells were constantly perfused with the slow flow of extracellular solution consisting of (mM): NaCl 140, KCl 3, $CaCl_2$ 2, $MgCl_2$ 1, HEPES 10 (Sigma-Aldrich Corp. Rehovot, Israel), supplemented with 2 mg/ml glucose (Sigma-Aldrich Corp. Rehovot, Israel), pH 7.3, osmolarity adjusted to 300-305 mOsm. The patch pipettes had resistances of 3–5 MOhm after filling with a solution containing (in mM): KCl 135, HEPES 10, glucose 5, MgATP 2, GTP 0.5 (Sigma-Aldrich Corp. Rehovot, Israel), pH 7.3, osmolarity adjusted to 285-290 mOsm. After obtaining the giga-ohm seal, the membrane was ruptured and the cells were subjected to fast current clamp by injecting an appropriate amount of current in order to adjust the membrane potential to about -70 mV. The experiments were taken into account only when this adjustment current was stable during the measurements. The changes in the neuronal membrane potential were acquired through a Digidata 1550 analog/digital converter using pClamp 10 electrophysiological software (Molecular Devices, Foster City, CA). The acquisition started upon receiving the TTL trigger from MEA setup. The signals were filtered at 10 kHz and digitized at 50 kHz. The cultures mainly consisted of pyramidal cells as a result of mainly enzymatic and mechanical dissociation. For patch clamp recordings, pyramidal neurons were intentionally selected based on their morphological properties.

MEA and Patch Clamp synchronization:

The experimental setup combines multi-electrode array, MEA 2100, and patch clamp. The multi-electrode array is controlled by the MEA interface board and a computer. The Patch clamp sub-system consists of several microstar manipulators, an upright microscope (Slicescope-pro-6000, Sceintifica), and a camera. Stimulations and recordings are implemented using multiclamp 700B and Digidata 1550A and are controlled by a second computer. The recorded MEA/patch data is saved on the computers respectively. The time of the MEA system is controlled by a clock placed in the MEA interface board and the time of the patch subsystem is controlled by a clock placed in the Digidata 1550A. The relative timings are controlled by triggers sent from the MEA interface board to the Digidata using leader-laggard configuration.

Extracellular electrode selection:

For the extracellular stimulations in the performed experiments an extracellular electrode out of the 60 electrodes was chosen by the following procedure. While recording intracellularly, all 60 extracellular electrodes were stimulated serially at 1-2 Hz and above-threshold, where each electrode is stimulated several times. The electrodes that evoked well-isolated, well-formed spikes were used in the experiments.

Extracellular threshold estimation:

After choosing an extracellular electrode, its threshold for stimulation was estimated. Stimulations at 0.5-1 Hz with duration in the range [200, 2000] μs and different values of voltage amplitudes were given, until a response failure occurred. The threshold was defined between the stimulation voltage that resulted in a response failure to the closest value of stimulation voltage that resulted in an evoked spike. For patched neurons that were significantly close to an extracellular electrode (several micrometers) shorter stimulation durations were used in order to avoid the stimulation artifact in the voltage recordings. The stability of the extracellular threshold was confirmed during the experiments. After the training of coupled intra- and extra-stimulations, the extracellular threshold was re-estimated in two methods; first, it was estimated 1 minute after training, second, it was estimated ~10 seconds after training with 2 predefined stimulation amplitudes. (Note that the MEA has three independent stimulators).

Intracellular threshold estimation:

In order to find a threshold for the intracellular stimulation, several stimulations at 1 Hz were given. The duration of the stimulations was set to 3 ms, and the intensity typically ranged from 100 pA and increased by 50 pA every stimulation until an evoked spike occurred.

Figure 1 experiments protocol:

An extracellular electrode was selected and both intra- and extra- cellular thresholds were estimated. The neuronal response latency, NRL, and its stability for the extracellular stimulations were estimated. The NRL was used to accurately adjust the time-lag between intracellular evoked spikes or EPSPs originated from consecutive intra- and extra- cellular stimulations to be in the range of 1-4 ms. We note that an above-threshold extracellular stimulation given shortly, e.g. 2 ms, after an above-threshold intracellular stimulation, does not result in an evoked spike, and can be used to enhance adaptation. The thresholds and NRL were rechecked at the end of the experiment, in order to ensure their stability.

Statistical analysis:

The demonstrated results were repeated tens of times on many cultures.

Data analysis:

Analyses were performed in a Matlab environment (MathWorks, Natwick, MA, USA). The recorded data from the MEA (voltage) was filtered by convolution with a Gaussian that had a standard deviation (STD) of 0.1 ms. Evoked spikes were detected by threshold crossing, typically -40 mV, using a detection window of [0.5, 30] ms following the beginning of an extracellular stimulation. In order to calculate the neuronal response latency, defined as the time-lag between the stimulation and its corresponding evoked spike, the evoked spikes' times were extracted from the recorded voltage.

**Simulations of biological neural networks:**

The perceptron:

The input layer consists of N input units and an output unit functioning as a leaky integrated and fire (LIF) neuron (see Output production). The input units are connected to the output unit via N synaptic weights, $W_m$ (Fig. 2b), or via K=N/5 dendritic strengths, $J_i$ (Fig. 2c). In the synaptic scenario, $\{W_m\}$ are the tunable parameters (Fig. 2b), whereas for the dendritic scenario, $\{J_i\}$ are the tunable parameters while $\{W_m\}$ are time-independent (Fig. 2c).

The supervised learning algorithm:

The scenario of supervised learning by a biological perceptron is examined using a teacher and a student. The mission of the student is to imitate the responses, i.e. the outputs, of the teacher, where both have the same architecture, hence it is a realizable rule. For each asynchronous input the teacher produces an output. The timings and the amplitudes of the inputs as well as the resulting teacher's firing timings are provided to the student. Those input/output relations constitute the entire information provided to the student for each asynchronous input. The algorithm is composed of 3 parts: Output production, weights adaptation and learning.

Output production: An identical asynchronous input, example, is given to the teacher and the student, each produces its outputs according to their weights and decaying input summation, $O^T$ and $O^S$, respectively (see Output production – Leaky integrate and fire neuron).

Weight adaptation: For each input unit the teacher preforms weights adaptation next to its output production, following its input/output (see Adaptation). The student preforms the same adaptation as the teacher, unless otherwise stated (see Student's adaptation).

Learning: The student preforms learning steps, unless otherwise stated, on weights with conflicting outputs with the teacher, i.e. $O_m^T \neq O_m^S$ for the $m^{th}$ input unit.

Inputs generation:

Each input was composed of N/2 randomly stimulated input units. For each stimulated unit a random delay and a stimulation amplitude were chosen from given distributions. The delays were randomly chosen from a uniform distribution with a resolution of 1 (0.001) ms, such that the average time-lag between two consecutive stimulations was 2 (10) ms for the synaptic (dendritic) scenario. Stimulation amplitudes were randomly chosen from a uniform distribution in the range [0.8, 1.2]. Note that the reported results are qualitatively robust to the scenario where all the non-zero amplitudes equal 1. In the dendritic scenario, the five $W_m$ connected to the same dendrite were stimulated sequentially in a random order and with an average time-lag of 10 ms between consecutive stimulations.

Output production – Leaky integrate and fire neuron:

In the synaptic adaptation scenario, the voltage of the output unit is described by the leaky integrate and fire (LIF) model

$$\frac{dv}{dt} = -\frac{v - v_{st}}{T} + \sum_{m=1}^{N} W_m \delta(t - \tau_m) \quad (1)$$

where v(t) is the scaled voltage, T = 20 ms is the membrane time constant and $v_{st} = 0$ stands for the scaled stable (resting) membrane potential. $W_m$ and $\tau_m$ stand for the $m^{th}$ weight and delay, respectively. A spike occurs when the voltage crosses the threshold, $v(t) \geq 1$ and at that time the output unit produces an output of 1, otherwise the output is 0. After a spike occurs, the voltage is set to $v_{st}$. For simplicity, we scaled the equation such that $v_{th} = 1$, $v_{st} = 0$, consequently, $v \geq 1$ is above threshold and $v < 1$ is below threshold. Nevertheless, results remain the same for both the scaled and unscaled equations, e.g. $v_{st} = -70$ mV and $v_{th} = -54$ mV. The initial voltage was set to $v_{(t=0)} = 0$.

In the dendritic adaptation scenario, the voltage of each dendritic terminal is described by

$$\frac{dv_i}{dt} = -\frac{v_i - v}{T} + J_i \cdot \sum_{m=\frac{N}{K}(i-1)+1}^{\frac{N}{K} \cdot i} W_m \delta(t - \tau_m) \tag{2}$$

where $v_i(t)$ and $J_i$ stand for the voltage and the strength of the $i^{th}$ dendrite, respectively. The rest of the parameters are identical to the synaptic adaptation scenario.

Adaptation:

The adaptation for the synaptic scenario is done according to

$$W^{t+1}{}_m = W^t_m \cdot (1 + \delta(\Delta t)) \tag{3}$$

where the discrete time t measures the number of asynchronous inputs and $\Delta t$ is the time-lag between a sub-threshold stimulation to $W_m$ (stimulation that didn't evoke spike, output 0) and an evoked spike, within one asynchronous input. Similarly, the dendritic adaptation is given by

$$J_i^{t+1} = J_i^t \cdot (1 + \delta(\Delta t)) \tag{4}$$

where $\Delta t$ now is the time-lag between a sub-threshold stimulation at $J_i$ and an evoked spike from another dendrite. For both scenarios

$$\delta(\Delta t) = A \cdot \exp\left(-\frac{\Delta t}{15}\right) \cdot sign(\Delta t) \tag{5}$$

representing the strengthening/weakening of a weight conditioned to a prior/later evoked spike at a time delay $\Delta t$, respectively, where a cutoff time window of 50 ms is enforced. For simplicity, a step function,

$$\delta(\Delta t) = \pm A \tag{6}$$

was used for all time delay $\Delta t$, unless otherwise stated. However, all results are robust to adaptation in the form of either exponential decay or a step function.

Student's adaptation:

In order to perform the same adaptation as the teacher, the required information is the teacher's temporal input/output relations. Note that although the student performs the same adaptation steps as the teacher, it does not necessarily ensure tracking of the parameters of the teacher, since the changes in the weights are relative to the current values of the weights of the student.

Learning:

Learning steps were performed on the student's weights with conflicting output with the teacher. This learning rule is based on a gradient descent dynamics, which minimizes a cost function

$$C = (v^T - v^S) \cdot (O^T - O^S)$$

that measures the deviation of the student voltage, $v^S$, form the teacher voltage, $v^T$, in case of an error (unmatched spike timings between the teacher and the student). A spike is considered as v=1. The change in the weights $W_m$ is proportional to the negative derivative of the cost function relative to the weight.

$$\Delta W_m \propto -\frac{dC}{dW_m} = -\frac{dC}{dv_m^S}\frac{dv_m^S}{dW_m}(O_m^T - O_m^S) = \frac{dv_m^S}{dW_m}(O_m^T - O_m^S) = x_m \exp\left(\frac{t-\tau_m}{T}\right)(O_m^T - O_m^S) \ .$$

$x_m$ denotes the stimulation amplitude of the m$^{th}$ input unit. For simplicity, the weighted exponential prefactor is neglected, where the qualitative results remain the same with/without the weighted exponential prefactor. Consequently, the learning step for the synaptic scenario is similar to the traditional perceptron learning algorithm

$$W_m^{t+1} = W_m^t + \eta \cdot (O_m^T - O_m^S) \cdot x_m \qquad (7)$$

and similarly for the dendritic scenario

$$J_i^{t+1} = J_i^t + \eta \cdot (O_m^T - O_m^S) \cdot x_m \ . \qquad (8)$$

where η denotes the learning step and $O_m^T$ and $O_m^S$ are the outputs of the teacher and the student at the m$^{th}$ input unit in the i$^{th}$ dendrite, respectively, and $x_m$ denotes the stimulation amplitude of the m$^{th}$ input unit.

Calculating the generalization error:

The estimation consisted of up to 20,000 inputs presented to the teacher and the student, where each input generates about 30/200 evoked spikes for the synaptic/dendritic scenario. The generalization error is defined as

$$\varepsilon_g = \frac{total\ no.\ of\ mismatch\ firing}{total\ no.\ of\ stimulations} \ . \qquad (9)$$

Figure 2:

Panel B: $\{W_m\}$ were chosen from a uniform distribution in the range [0.1, 0.2]. Adaptation was following equation (6) with A = 0.05, and learning was following equation (7) with η = 1/1000.

$W_m$ was bounded from above by 1.5 and from below by $10^{-4}$. The fixed learning rate was compared to the accelerating method using adaptive learning step:

$$\eta^{t+1} = \eta^t \cdot e^{-\tau} + B \cdot sign(O^T - O^S)$$

using $\tau = 0.1$, B = 0.01 and $\eta$ was initiated as 1/1000.

Panel C: $\{W_m\}$ were chosen from a uniform distribution in the range [0.1, 0.9] and then were normalized to a mean equals to 0.5. $\{J_i\}$ were chosen from a uniform distribution in the range [0.5, 1.5]. Stimulations with low amplitudes (0.01) were given to the N/2 unstimulated input units, resulting in non-frozen $J_i$. The adaptation follows equation (5) with A=0.05 and the learning follows equation (7) with η =1/1000. $J_i$ was bounded from below by 0.1 and from above by 3. The fixed learning rate was compared to the accelerating method using adaptive learning step:

$$\eta^{t+1} = \eta^t \cdot e^{-\tau} + B \cdot sign(O^T - O^S)$$

using $\tau = 0.1$, B = 0.01 and $\eta$ was initiated as 1/1000.

**Simulations of neural network:**

Architecture:

The feedforward neural network contains 784 input units, 30 hidden units and 10 output units in a fully connected architecture. Each unit in the hidden and the output layers has an additional input from a bias unit. Weights from the input layer to the hidden layer, $W_1$, and from the hidden layer to the output layer, $W_2$, were randomly chosen from a Gaussian distribution with a zero average and standard deviation equals 1. All weights were normalized such that all input weights to each hidden unit have an average equals 0 and a STD equals 1. The initial value of the bias of each weights was set to 1. We trained the network on the handwritten digits dataset, MNIST, using gradient descent. The inputs, examples from the train dataset, contain 784 pixel values in the range [0, 255]. We normalized the inputs such that the average and the STD are equal to 0 and 1, respectively.

Forward propagation:

The output of a single unit in the hidden layer, $a_j^1$, was calculated as:

$$z_j^1 = \sum_i (W_{ij}^1 \cdot X_i) + b_j^1$$

$$a_j^1 = \frac{1}{1 + e^{-z_j^1}}$$

where $W_{ij}^1$ is the weight from the i[th] input to the j[th] hidden unit, $X_i$ is the i[th] input and $b_j^1$ is the bias to the j[th] hidden unit.

For the output layer, the output of a single unit, $a_j^2$ was calculated as:

$$z_j^2 = \sum_i (W_{ij}^2 \cdot a_i^1) + b_j^2$$

$$a_j^2 = \frac{1}{1 + e^{-z_j^2}}$$

where $W_{ij}^2$ is the weight from the i[th] hidden unit to the j[th] output unit, $X_i$ is the i[th] input and $b_j^2$ is the bias to the j[th] output unit.

Back propagation:

We used two different cost functions; the first was the cross entropy:

$$C = -\frac{1}{N} \sum_n [y * \log(a) + (1 - y) * \log(1 - a)]$$

and the second was the mean square error (MSE):

$$C = \frac{1}{2N} \sum_n (y - a)^2$$

where y are the desired labels and $a$ stands for the current 10 output units of the output layer. The summation is over all training examples, N.

The backpropagation method computes the gradient for each weight with respect to the chosen cost function. The weights and biases were updated according to 3 different methods:

(1) <u>Momentum:</u>

The weights update:

$$W^{t+1} = (1 - \alpha) \cdot W^t + V^{t+1}$$
$$V^{t+1} = \mu \cdot V^t - \eta_0 \cdot \nabla_{W^t} C$$

where t is the discrete time-step W are the weights, $\alpha$ is a regularization constant, $\eta$ is the fixed learning rate, and $\nabla_{W^t} C$ is the gradient of the cost function for each weight at time t. V was initialized as: $-\eta_0 \cdot \nabla_W C_{first}$, where $\nabla_W C_{first}$ is the first computed gradient and the biases update:

$$V_b^{t+1} = \mu \cdot V_b^t - \eta_0 \cdot \nabla_{b^t} C$$

$$b^{t+1} = b^t + V_b^{t+1}$$

where $\nabla_b C$ is the gradient of the cost function of each bias with respect to its weight, b, and $V_b$ was initialized as: $-\eta \cdot \nabla_b C_{first}$, where $\nabla_b C_{first}$ is the first computed bias gradient.

(2) <u>Acceleration:</u>

The weights update:

$$W^{t+1} = (1 - \alpha) \cdot W^t - |\eta^{t+1}| \cdot \nabla_{W^t} C$$

$$\eta^{t+1} = \eta^t \cdot e^{-\tau} + A_{1/2} \cdot tanh(\beta_{1/2} \cdot \nabla_{W^t} C)$$

where $\eta$ is defined for each weight, $A_1$ and $\beta_1$ are constants representing the amplitude and the gain between the input and the hidden layers, respectively, and $A_2$ and $\beta_2$ represent the same between the hidden and the output layers. $\eta$ was initialized as: $A_{1/2} \cdot tanh(\beta_{1/2} \cdot \nabla_W C_{first})$, where $\nabla_W C_{first}$ is the first computed gradient and the biases update:

$$b^{t+1} = b^t - |\eta_b^{t+1}| \cdot \nabla_{b^t} C$$

$$\eta_b^{t+1} = \eta_b^t \cdot e^{-\tau} + A_{1/2} \cdot tanh(\beta_{1/2} \cdot \nabla_{b^t} C)$$

(3) <u>Advanced acceleration:</u>

The weights update:

$$W^{t+1} = (1 - \alpha) \cdot W^t + V^{t+1}$$

$$V^{t+1} = \mu \cdot V^t - \eta^{t+1} \cdot \nabla_{W^t} C$$

$$\eta^{t+1} = \eta^t \cdot e^{-\tau} + A_{1/2} \cdot tanh(\beta_{1/2} \cdot \nabla_{W^t} C)$$

and the biases update:

$$b^{t+1} = b^t + V_b^{t+1}$$

$$V_b^{t+1} = \mu \cdot V_b^t - \eta_b^{t+1} \cdot \nabla_{b^t} C$$

$$\eta_b^{t+1} = \eta_b^t \cdot e^{-\tau} + A_{1/2} \cdot tanh(\beta_{1/2} \cdot \nabla_{b^t} C)$$

Testing the network:

The network classification accuracy was tested on the MNIST dataset for testing, containing 10,000 inputs. The test inputs were also normalized to have an average of each equals to 0 and a STD equals to 1.

Optimization:

For each update method the parameters were chosen to maximize the test accuracy. For optimization we first used a grid of the adjustable parameters followed by a fine tuning with higher resolution for each parameter. The optimization was performed over 3 parameters for the momentum method $(\mu, \eta_0, \alpha)$, 6 parameters for the acceleration method $(A_1, A_2, \beta_1, \beta_2, \tau, \alpha)$ and for 7 parameters for the advanced acceleration method $(A_1, A_2, \beta_1, \beta_2, \tau, \alpha, \mu)$.

Figure 3:

Panel B: The feed forward neural network was presented with 300 examples with equally number of appearance of each digit. The cross entropy cost function was used and the following parameters for each method: momentum with $\mu = 0.9, \eta_0 = 0.02, \alpha = 0$, acceleration with $A_1 = 0.1, A_2 = 0.27, \beta_1 = 5000, \beta_2 = 900, \tau = 0.598, \alpha = 0.003$, and advanced acceleration with $A_1 = 0.07, A_2 = 0.07, \beta_1 = \infty, \beta_2 = \infty, \tau = 0.29, \alpha = 0.005, \mu = 0.5$. Results are presented as the average of 100 different runs, and typical error bars are presented for the last point.

Panel C: The feed forward neural network was presented with 6000 examples randomly taken from the 60000 examples in the training dataset, and presented to the network with 30 mini-batches of size 200. The cross entropy cost function was used and the following parameters for each method: momentum with $\mu = 0.68, \eta_0 = 0.65, \alpha = 0.08$, acceleration with $A_1 = 3, A_2 = 0.7, \beta_1 = 1500, \beta_2 = 1000, \tau = 0.4, \alpha = 0.065$, and advanced acceleration with $A_1 = 0.5, A_2 = 0.5, \beta_1 = \infty, \beta_2 = \infty, \tau = 0.1, \alpha = 0.1, \mu = 0.55$. Results are presented as the average of 100 different runs, and typical error bars are presented for the last point.

Panel D: The feed forward neural network was presented with 1200 examples randomly taken from the 60000 examples in the training dataset, and presented to the network with 24 mini-batches of size 50. The cross entropy cost function was used and the following parameters for each method: momentum with $\mu = 0.75, \eta_0 = 0.6, \alpha = 0.11$, acceleration with $A_1 = 1.15, A_2 = 0.6, \beta_1 = 4500, \beta_2 = 3500, \tau = 0.1, \alpha = 0.055$, and advanced acceleration with $A_1 = 0.55, A_2 = 0.55, \beta_1 = \infty, \beta_2 = \infty, \tau = 0.1, \alpha = 0.1, \mu = 0.55$. Results are presented as the average of 100 different runs, and typical error bars are presented for the last point.

Panel E: The feed forward neural network was presented with 60 with equally number of appearance of each digit. The 60 examples were presented to the network 5 times. The cross entropy cost function was used and the following parameters for each method: momentum with $\mu = 0.87, \eta_0 = 0.035, \alpha = 0.005$, acceleration with $A_1 = 0.11, A_2 = 0.26, \beta_1 = 2000, \beta_2 = 2500, \tau = 0.45, \alpha = 0.008$, and advanced acceleration with $A_1 = 0.035, A_2 = 0.02, \beta_1 = $

$4500, \beta_2 = 5, \tau = 0.0619, \alpha = 0.01, \mu = 0.6$. Results are presented as the average of 100 different runs, and typical error bars are presented for the last point.

Panel F: The feed forward neural network was presented with 300 examples with equally number of appearance of each digit. The mean-square-error cost function was used and the following parameters for each method: momentum with $\mu = 0.6, \eta_0 = 0.35, \alpha = 0.005$, acceleration with $A_1 = 0.95, A_2 = 0.25, \beta_1 = 5000, \beta_2 = 40, \tau = 0.15, \alpha = 0.008$, and advanced acceleration with $A_1 = 0.06, A_2 = 0.09, \beta_1 = 2100, \beta_2 = 1, \tau = 0.015, \alpha = 0.005, \mu = 0.8$. Results are presented as the average of 100 different runs, and typical error bars are presented for the last point.

Supplementary Figure 3:

The feed forward neural network was presented with 300 examples with equally number of appearance of each digit. The examples are composed of 5 subsets of 60 balanced examples, where in every subset each label appears exactly 6 times. The cross entropy cost function was used and the following parameters for each method: momentum with $\mu = 0.87, \eta_0 = 0.035, \alpha = 0.005$, acceleration with $A_1 = 0.11, A_2 = 0.26, \beta_1 = 2000, \beta_2 = 2500, \tau = 0.45, \alpha = 0.008$, and advanced acceleration with $A_1 = 0.035, A_2 = 0.02, \beta_1 = 4500, \beta_2 = 5, \tau = 0.0619, \alpha = 0.01, \mu = 0.6$. Results are presented as the average of 100 different runs, and typical error bars are presented for the last point.

Supplementary Figure 4:

The feedforward neural network was presented with 300 examples with equally number of appearance of each digit. The examples are composed of 30 subsets of 10 balanced examples, where in every subset each label appears exactly one time. Results are also presented for the advanced acceleration method where for each subset of 10 examples, the labels were in a fixed order. The cross entropy cost function was used and the following parameters for each method: momentum with $\mu = 0.87, \eta_0 = 0.035, \alpha = 0.005$, acceleration with $A_1 = 0.11, A_2 = 0.26, \beta_1 = 2000, \beta_2 = 2500, \tau = 0.45, \alpha = 0.008$, and advanced acceleration with $A_1 = 0.035, A_2 = 0.02, \beta_1 = 4500, \beta_2 = 5, \tau = 0.0619, \alpha = 0.01, \mu = 0.6$. Results are presented as the average of 100 different runs, and typical error bars are presented for the last point.

**Acknowledgments**

We would like to thank Editage (www.editage.com) for English language editing.

**Author Contributions**

S.S. and R.V. contributed equally to all experimental result aspects while Y.T. prepared the tissue cultures and materials for the experiments. S.S. and Y.M. contributed equally to the simulation results. S.H. confirmed some of the simulation results. A.G. contributed to conceptualization and for the data analysis. I.K. initiated the study and supervised all aspects of the work. All authors discussed the results and commented on the manuscript.

**Competing Interests**

Authors declare no competing interests.

**Additional information**

Supplementary information is available for this paper at

Correspondence and requests for materials should be addressed to I.K.

# Figures

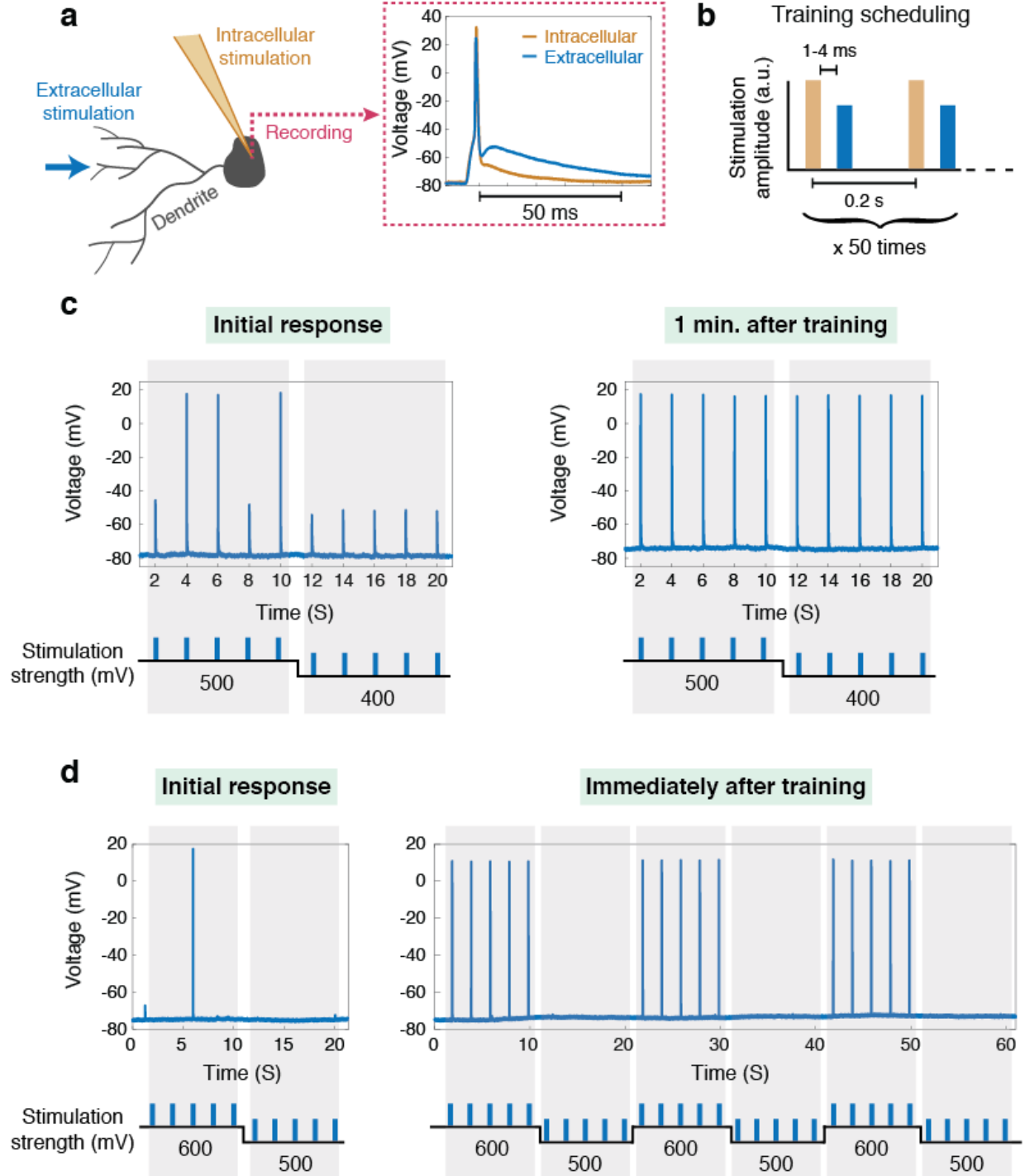

**Figure 1.** Experimental results indicate that adaptation rates increase with training frequency. (**a**) The experimental scheme where a patched neuron is stimulated intracellularly via its dendrites (Materials and Methods) and a different spike waveform is generated for each stimulated route.

(**b**) The scheduling for coherent training consists of repeated pairs of intracellular stimulations (orange) generating a spike followed by an extracellular stimulation (blue) with the lack of a spike. (**c**) An example of the first type of experiments, where decreasing extracellular stimulation amplitude, is used to estimate the threshold using intracellular recording (left), and enhanced responses measured a minute after the termination of the training, **b** (right). (**d**) An example of the second type of experiment, similar to **c**, where enhanced responses are observed 10 seconds after the termination of the training (Materials and Methods).

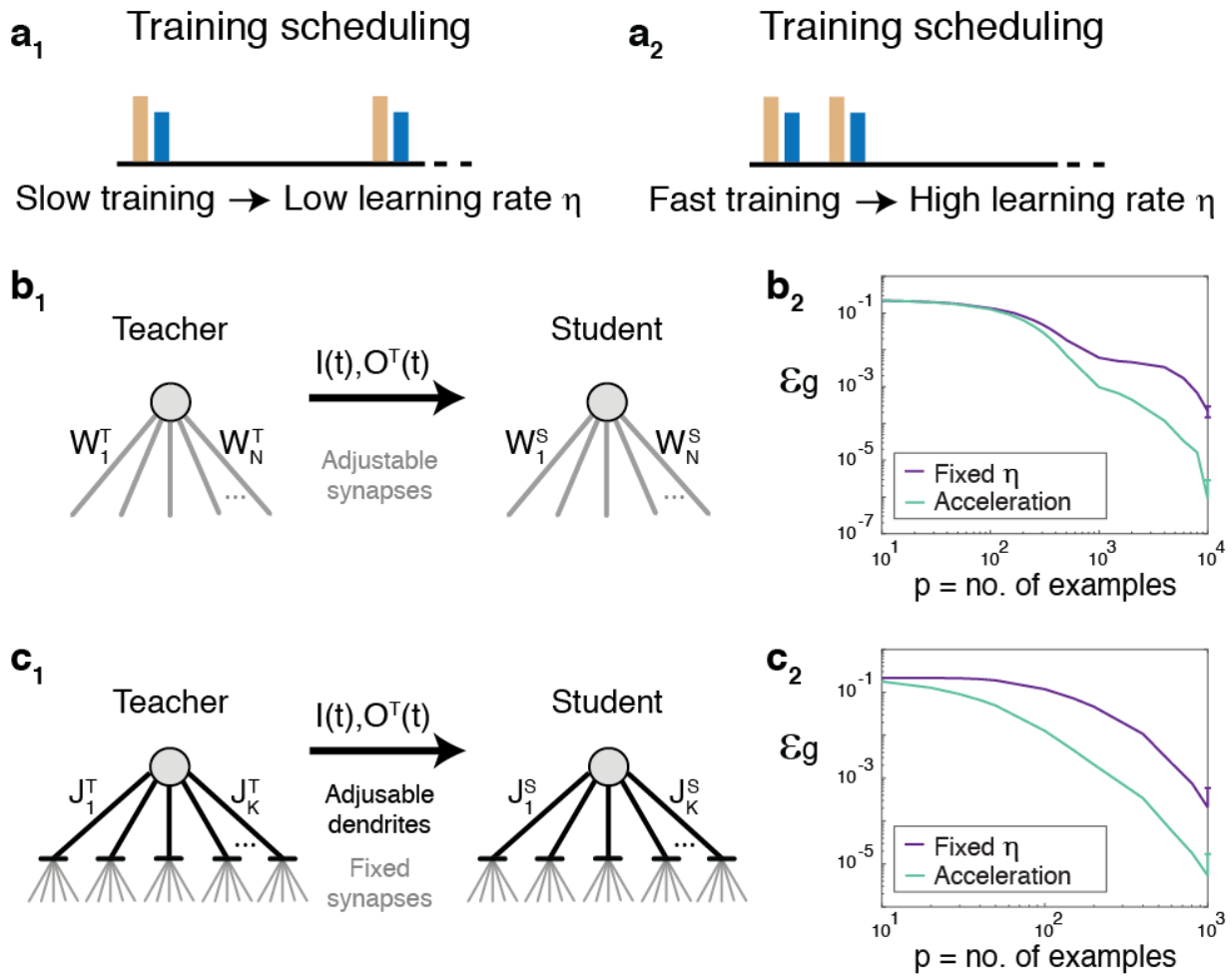

**Figure 2.** Acceleration of supervised realizable learning rules based on the biologically inspired mechanism. (**a**) The implication of the biological mechanism (Fig. 1) indicating that training scheduling with low/high frequency (**a₁**/**a₂**) results in a low/high learning rate, $\eta$. (**b**) Synaptic learning. (**b₁**) A perceptron with 1000 asynchronous inputs and a leaky integrate-and-fire output

unit. The synapses of the teacher/student, $W^T/W^S$, are dynamically updated following a biological adaptation rule, a spike-time-dependent-plasticity (Materials and Methods). (**b₂**) The generalization error, $\varepsilon_g$, for the training of the student in **b₁** using fixed learning step $\eta = 0.001$ (purple) and for the accelerating scenario $\eta^{t+1} = \eta^t \cdot \exp(-0.1) + 0.01 \cdot sign(O^T - O^S)$ (green), where $O^T/O^S$ stands for a teacher/student output, and the initial $\eta = 0.001$. (**c**) Dendritic learning. (**c₁**) Dendritic learning, similar to **b₁**, where the 1000 synapses are fixed and the 200 dendrites of the teacher/student, $J^T/J^S$, are updated using similarly adaptive rule as in **b** (Materials and Methods). (**c₂**) $\varepsilon_g$ for the trained student, **c₁**, using the same scenarios for $\eta$ as in **b₂**. Results, **b₂** and **c₂**, are averaged over 30 training datasets and the STD of $\varepsilon_g$ is presented for the maximal presented p.

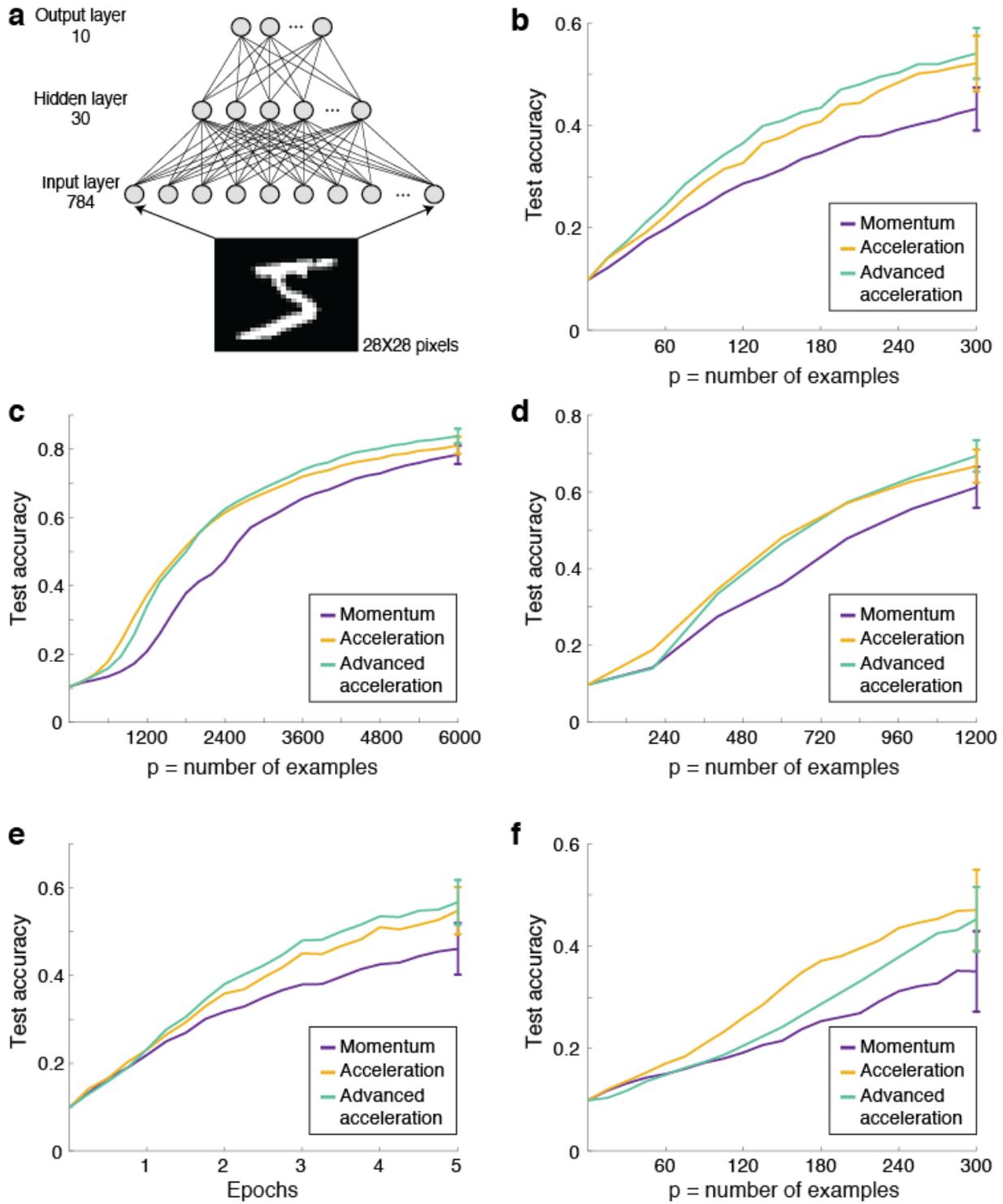

**Figure 3.** Biological-inspired accelerating learning for the MNIST database in comparison with a common existing learning method. (**a**) The trained network, using backpropagation and cross-

entropy cost function, consists of 784 (28 x 28) input units representing the pixels of an MNIST digit, 30 hidden units, and 10 outputs representing the probabilities for the possible labels (Materials and Methods). (**b**) The maximal attainable test accuracy for 300 examples trained only once by the network in **a**, using the following methods; momentum, eq. (3) (purple), acceleration, eq. (4) (orange) and advanced acceleration, eq. (5) (green). (**c**) Similar to **b** using 6000 examples composed of 30 batches of 200. (**d**) Similar to **b** using 1200 examples composed of 24 batches of 50. (**e**) 60 examples are trained 5 times. The total number of presented examples to the network, **a**, is 300 (60 x 5) as in **b**. (**f**) Similar to **b** using mean-square-error cost-function. Results, panels **b-f**, are averaged over 100 training datasets and the STD of the test accuracy is presented for the last trained example.

# Supplementary Information

# Brain experiments imply adaptation mechanisms which outperform common AI learning algorithms

Shira Sardi[1], Roni Vardi[2], Yuval Meir[1], Yael Tugendhaft[1], Shiri Hodassman[1], Amir Goldental[1] & Ido Kanter[1,2]
[1]Department of Physics, Bar-Ilan University, Ramat-Gan, 52900, Israel.
[2]Gonda Interdisciplinary Brain Research Center and the Goodman Faculty of Life Sciences, Bar-Ilan University, Ramat-Gan, 52900, Israel.

Correspondence to: ido.kanter@biu.ac.il

**Supplementary Figures**

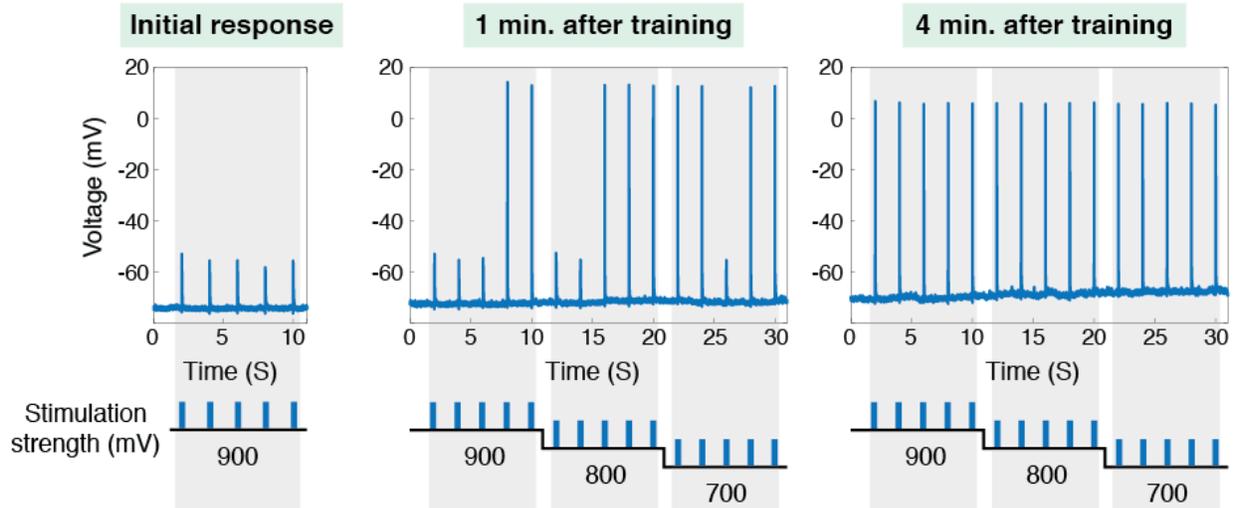

**Supplementary Figure 1. The effect of adaptation after one minute is further enhanced after additional several minutes.**

An example of the first type of experiments, as in Fig. 1c, where decreasing extracellular stimulation amplitude is used to estimate the threshold using intracellular recording (left), and enhanced responses measured one minute after the termination of the training, (middle). Further enhancement in the response is measured 4 minutes after termination of the training (right).

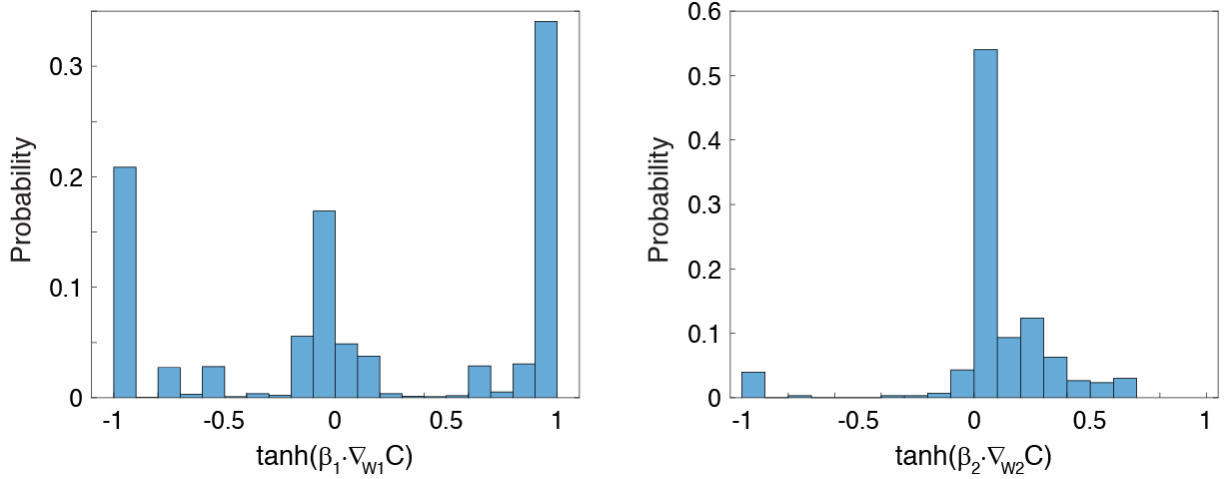

**Supplementary Figure 2. The maximal test accuracy for the advanced acceleration method is achieved in the finite $\beta$ limit.**

Left: The probability distribution for $\tanh(\beta_1 \cdot \nabla_{W_1} C)$, taken from the training of last example in Fig. 3e. Right: The probability distribution for $\tanh(\beta_2 \cdot \nabla_{W_2} C)$, taken from the training of last example in Fig. 3e. Both histograms indicate that the maximal test accuracy is obtained using finite $\beta$, i.e. $|\tanh(\beta_{1/2} \cdot \nabla_{W_{1/2}} C)| < 1$ with finite probability, which cannot be simplified to $sign(\nabla_{W_{1/2}} C)$.

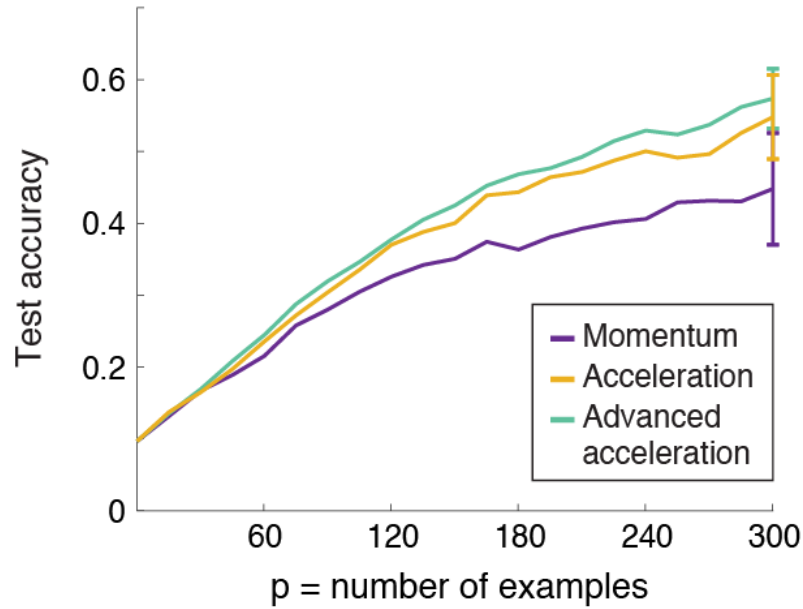

**Supplementary Figure 3. Maximal test accuracy for training set composed of 5 balanced subsets of 60 examples.**

The maximal attainable test accuracy for 300 examples trained only once by the network in Fig. 3a. The training set of 300 distinct examples was composed of 5 subsets of 60 balanced examples, i.e. each one of the labels appears 6 times, using the following methods; momentum, eq. (3) (purple), acceleration, eq. (4) (orange) and advanced acceleration, eq. (5) (green), and the test accuracy is ~0.57. The parameters are given in the Methods and Materials section. Results were averaged over 100 training sets and the STD is presented for p=300.

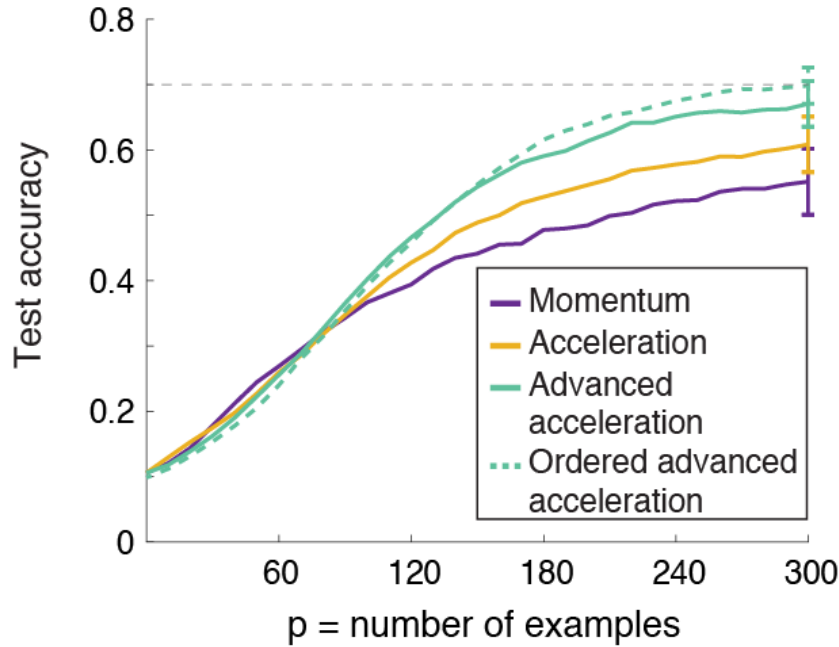

**Supplementary Figure 4. Maximal test accuracy for a training set composed of 30 subsets of 10 examples, where each label appears once.**

The maximal attainable test accuracy for 300 examples trained only once by the network in Fig. 3a. The training set of 300 distinct examples was composed of 30 subsets of 10 examples, where each label appears once, using the following methods; momentum, eq. (3) (purple), acceleration, eq. (4) (orange) and advanced acceleration, eq. (5) (green), and the test accuracy is ~0.67. The advanced acceleration with a fixed order of labels within all 30 subsets of 10 examples results in a test accuracy of ~0.7 (dashed green). The parameters are given in the Methods and Materials section. Results were averaged over 100 training sets and the STD is presented for p=300.